\documentclass{emulateapj}

\def\gtsim{
\mathrel{\raise.3ex\hbox{$>$}\mkern-14mu\lower0.6ex\hbox{$\sim$}}
}
\def\ltsim{
\mathrel{\raise.3ex\hbox{$<$}\mkern-14mu\lower0.6ex\hbox{$\sim$}}
}
\def\farcs{\hbox{$.\!\!^{\prime\prime}$}}
\def\deg{\hbox{$^\circ$}}
\def\fs{\hbox{$.\!\!^{\rm s}$}}

\slugcomment{Submitted to the Astrophysical Journal}

\shorttitle{Expansion Asymmetry and Age of Cas A}
\shortauthors{Fesen et al. }
\begin{document}

\title{The Expansion Asymmetry and Age of the Cassiopeia A Supernova Remnant\altaffilmark{1} } 

\author{Robert A.\ Fesen\altaffilmark{2}, 
        Molly C. Hammell\altaffilmark{2}, 
        Jon Morse\altaffilmark{3},
        Roger A.\ Chevalier\altaffilmark{4},
        Kazimierz J.\ Borkowski\altaffilmark{5},
        Michael A.\ Dopita\altaffilmark{6},
        Christopher L.\ Gerardy\altaffilmark{7},
        Stephen S.\ Lawrence\altaffilmark{8},
        John C.\ Raymond\altaffilmark{9}, \&
        Sidney van den Bergh\altaffilmark{10}
                                            } 
\altaffiltext{1}{Based on observations with the NASA/ESA Hubble Space Telescope,
obtained at the Space Telescope Science Institute,
which is operated by the Association of Universities for Research in
Astronomy, Inc.\  under NASA contract No.\ NAS5-26555.}
\altaffiltext{2}{6127 Wilder Lab, Department of Physics \& Astronomy, Dartmouth College, Hanover, NH 03755} 
\altaffiltext{3}{Department of Physics and Astronomy, Arizona State University, Box 871504, Tempe, AZ 85287}
\altaffiltext{4}{Department of Astronomy, University of Virginia, P.O. Box 3818, Charlottesville, VA 22903  }
\altaffiltext{5}{Department of Physics, North Carolina State University, Raleigh, NC 27695 }
\altaffiltext{6}{Research School of Astronomy and Astrophysics, The Australian National University,
                 Cotter Road, Weston Creek, ACT 2611 Australia}
\altaffiltext{7}{Astrophysics Group, Imperial College London, Blackett Laboratory, Prince Consort Road, London SW7 2BZ}
\altaffiltext{8}{Department of Physics and Astronomy, Hofstra University, Hempstead, NY 11549}
\altaffiltext{9}{Harvard-Smithsonian Center for Astrophysics, 60 Garden Street, Cambridge, MA 02138}
\altaffiltext{10}{Dominion Astrophysical Observatory, Herzberg Institute of Astrophysics, NRC of Canada,
                 5071 West Saanich Road, Victoria, BC V9E 2E7, Canada  }

\begin{abstract}

{\sl HST} images of the young supernova remnant Cas~A are used to explore the
expansion and spatial distribution of its highest velocity debris.  ACS/WFC
images taken in March and December $2004$ with Sloan F625W, F775W, and F850LP
filters were used to identify 1825 high-velocity, outlying ejecta knots through
measured proper motions of $0\farcs35 - 0\farcs90$ yr$^{-1}$ corresponding to
V$_{\rm trans} = 5500 - 14500$ km s$^{-1}$ assuming d = 3.4 kpc.  The
distribution of derived transverse expansion velocities for these ejecta knots
shows a striking bipolar asymmetry with the highest velocity knots (V$_{\rm
trans}$ $\geq$ 10500 km s$^{-1}$) confined to nearly opposing northeast and
southwest `jets' at P.A. = $45{\deg} - 70{\deg}$ and $230{\deg} - 270{\deg}$,
respectively. The jets have about the same maximum expansion velocity of
$\simeq$14000 km s$^{-1}$ and appear kinematically and chemically distinct in
that they are the remnant's only S-rich ejecta with expansion velocities above
the $10000 - 11000$ km s$^{-1}$ exhibited by outer nitrogen-rich ejecta which
otherwise represent the remnant's highest velocity debris.  In addition, we
find significant gaps in the spatial distribution of outlying ejecta in
directions which are  approximately perpendicular to the jets  (P.A. =
$145{\deg} - 200{\deg}$ and $335{\deg} - 350{\deg}$).  The remnant's central
X-ray point source lies some $7''$ to the southeast of the estimated expansion
center (PA = $169{\deg} \pm 8.4{\deg}$) indicating a projected motion toward
the middle of the broad southern ejecta knot gap.  Extrapolations of measured
nine month proper motions for all 1825 outer ejecta knots and a selected
subsample of 72 bright and compact knots suggest explosion dates (assuming no
knot deceleration) of $1662 \pm 27$ and $1672 \pm 18$, respectively.  We find
some evidence for non-uniform deceleration in different directions around the
remnant and find 126 knots located along the northwestern limb among the least
decelerated ejecta suggesting a convergence date of $1681 \pm19$.  A remnant
age of around 325 yr would imply a $\simeq$ 350 km s$^{-1}$ transverse
velocity for the central X-ray point source.
 
\end{abstract}


\section{Introduction}

The relative roles of neutrino heating and bipolar MHD jets as the underlying
mechanism behind core-collapse supernovae (SNe) are controversial
\citep{Janka03,Wheeler03}.  However, despite current uncertainties about the
specific engine that drives core-collapse explosions, a variety of observations
and hydrodynamic modeling make a compelling case that high-mass SNe are
intrinsically aspherical events.

Observations of extragalactic core-collapse SNe show increasing late-time
linear polarization levels suggesting that the innermost layers driving the SN
expansion are aspherical
\citep{Trammell93,Wang96,Wang01,Leonard00,Leonard01,LF01}.  In the case of
SN~1987A, spectropolarimetric observations and the early detection of
gamma-rays and hard X-rays indicating the transport of freshly synthesized
$^{56}$Ni from the core to the H-rich envelope lead to a model of envelope
asymmetries and ejecta fragmentation \citep{Arnett89,Chevalier89}.  Recent {\it
Hubble Space Telescope} ({\sl HST}) images of SN~1987A reveal an elongated, axially
symmetric remnant \citep{Wang02}.

State-of-the-art numerical simulations also show that spherically symmetric
core-collapses do not yield successful explosions
\citep{Rampp00,Lie01,Lie05,Buras03,Thompson03}. This has lead some modelers to
investigate the effects of rapid rotation and magnetic fields leading to
magnetorotational jet models \citep{Symbal84,Khok99,Wheeler00,Hoflich99,Hoflich01,Akiyama03}.
Others have investigated asymmetric
neutrino-driven models in order to generate aspherical sometimes even jet-like SN explosions
\citep{Burrows95,Shimizu01,Kifonidis03,Mad04,Janka05,Yamasaki05,Wilson05}.
Asymmetries in either the neutrino heating or MHD jets might also explain pulsar `kick'
velocities \citep{Fryer04,Scheck04,Kotake05}.

Observations of `long-duration' gamma-ray bursts (GRBs) suggest highly
aspherical core-collapse explosions.  In the popular `collapsar' GRB model, a
high mass core-collapse SN creates a black hole generating bipolar relativistic
jets \citep{Woosley93,MW99,MWH01}. These jets excite an external shock from
which energetic electrons, created through interaction with the ambient medium,
radiate synchrotron emission giving rise to broadband afterglows.  Direct
connections between SNe and GRBs include the observational coincidences between
GRBs and SN~1998bw and SN~2003dh, `bumps' in some GRB afterglow light-curves
consistent with underlying SN explosions
\citep{Galama98,Hjorth03,Stanek03,Matheson03,Kawa03,Mal04,Mazzali05}, and indications
that GRBs are related to star forming regions \citep{Pac98,Fruchter99}.  

Compared to SN observations, evidence for aspherical SN explosions based on
supernova remnant observations is much less clear. One of the most often cited
asymmetrically expanding SNRs is that of Cassiopeia A (Cas A), currently the
youngest known ($\sim$300 yr) Galactic, core-collapse remnant.  On even the
earliest photographic plate images, a `flare' or `jet' of knots and filaments
could be seen extending out along the northeast limb about $3.8'$ from the
remnant center at a PA $\sim$ 70${\deg}$ \citep{min68,vdbD70}.  Whereas
the main $\simeq$2$'$ radius emission ring of ejecta expands at velocities
$4000 - 6000$ km s$^{-1}$, debris in the NE jet have estimated velocities more
than twice as large, up to 14000 km s$^{-1}$ \citep{Fes96,Fesen01}.  However,
both the main shell's so-called `Fast-Moving Knots' (FMKs) and NE jet ejecta knots
have similar emission line spectra; namely, strong lines of [S~II]
$\lambda\lambda$6716,6731, suggesting similar chemical make-ups.  A southwest
`counterjet' of high-velocity, [S~II] emitting ejecta was recently discovered
\citep{Fesen01} and confirmed in X-rays and the infrared
\citep{Hwang04,Krause05}.

While the presence of a jet and counterjet in this high-mass progenitor SNR
might indicate an aspherical SN expansion, the nature of these jet features is
controversial. Proposed explanations include uneven ejecta decelerations due to
local ISM density variations and cavities \citep{min68,Blondin96,Blondin01},
bipolar MHD jets \citep{Khok99,Wheeler02}, or mildly asymmetrical
neutrino-driven explosions \citep{Janka05,Burrows05}. Complicating matters,
the NE and SW jets are not the only outlying ejecta around Cas~A, with several
dozen high-velocity, nitrogen-rich knots located in other regions of the
remnant \citep{Fes87,Fesen01}.

In this paper, we report results from a deep {\sl HST} imaging survey of the
Cas~A SNR.  These images reveal a large population of high-velocity knots of
ejecta with an asymmetric distribution.  A complete catalog of outer
ejecta knots will be presented in a separate paper \citep{HF06}. Here we
address the overall spatial distribution and expansion velocity of the
remnant's outer ejecta and discuss limits on the dynamical age of the remnant.
The observations and knot flux measurement procedures are described in $\S$2
and $\S$3, with the results presented and discussed in $\S$4 and $\S$5,
respectively.

\section{Observations}

High resolution {\sl HST} images of the Cas~A remnant were obtained on 4--6
March 2004 and 4--5 December 2004 using the Wide Field Channel (WFC) of the
Advanced Camera for Surveys (ACS; \citealt{Ford98,Pavlovsky04}).
The ACS/WFC consists of two $2048 \times 4096$ CCDs with an average
image pixel scale of $0\farcs05$ providing a $202'' \times 202''$ field of view.  Four
2-point line dithered images were taken in each of the four ACS/WFC Sloan
Digital Sky Survey (SDSS) filters, namely F450W, F625W, F775W, and F850LP
(i.e., SDSS g,r,i, and z), at each target position to permit cosmic ray
removal, coverage of the $2\farcs5$ interchip gap, and to minimize saturation
effects of bright stars in the target fields.

Total integration times in the F450W, F625W, F775W, and F850LP filters were
2000 s, 2400 s, 2000 s, and 2000 s, respectively.  Standard pipeline
IRAF/STSDAS\footnote{IRAF is distributed by the National Optical Astronomy
Observatories, which is operated by the Association of Universities for
Research in Astronomy, Inc.\ (AURA) under cooperative agreement with the
National Science Foundation. The Space Telescope Science Data Analysis System
(STSDAS) is distributed by the Space Telescope Science Institute.} data
reduction was done including debiasing, flat-fielding, geometric distortion
corrections, photometric calibrations, and cosmic ray and hot pixel removal.
The STSDAS {\it drizzle} task was used to combine exposures in
each filter.  Due to significant reddening toward Cas A (A$_{\rm V}$ = $4.5 -
8$ mag; \citealt{Hur96,Reynoso02}), [O~III] $\lambda\lambda$4959,5007 line
emission was too weak to be detected for most outlying knots and we have not
included F450W images in our analysis. 

We measured outer knot ACS/WFC fluxes from the three remaining SDSS filter
image sets using the automated source extraction software package SExtractor
\citep{Bertin96}.  In cases where the SExtractor program failed to return a
reasonable flux or failed to return a flux at all, the knot fluxes were
calculated by hand.  In all cases, the fluxes were calculated using 5 pixel
apertures.  Background estimates were performed by SExtractor using a $24$
pixel rectangular annulus about the isophotal limits of the object.  When
fluxes were calculated manually, background estimation was performed by
calculating the total $5$ pixel aperture flux in at least five positions near
the object (avoiding other sources) and then subtracting the mean computed
``background'' flux from the total object pixel sum.  Most knots whose fluxes
required manual computation were located near a bright background source or
very close to another ejecta knot.

\section{Outer Knot Identification and Fluxes}

High-velocity, outer ejecta knots were identified through proper motion
measurements on the March and December 2004 ACS/WFC images \citep{HF06}.
Ejecta knots were defined as being high-velocity if their epoch 2004.3 radial
distance exceeded 100$''$ from the remnant's center of expansion (V$_{\rm exp}
\sim$ 5500 km s$^{-1}$) and they showed a proper motion $\geq$ $0\farcs35$ yr$^{-1}$
(see \citealt{HF06}).

Following \citet{HF06} and \citet{Fesen06}, we used the filter fluxes to bin
outlying ejecta knots into three emission classes; namely, strong [N~II]
$\lambda\lambda$6548,6583 emission knots, strong [O~II]
$\lambda\lambda$7319,7330 emission knots, and strong [S~II]
$\lambda\lambda$6716,6731 FMK-like knots. Flux ratio criteria between these
three classes were chosen to segregate knots with similar ratios seen for main
shell or outer ejecta knots with existing spectroscopic data. For example,
outlying ejecta knots whose $5000 - 7500$ \AA\ spectra show largely just
[N~II] $\lambda\lambda$6548,6583 emission like those discussed by
\citet{Fesen01} exhibit F625W/F775W and F625W/F850LP ratios more than an order
of magnitude larger than [S~II] bright knots. In similar fashion, the newly
discovered outlying O-rich knots which show a 6000 - 7500 \AA\ spectrum with
strong [O~I] and [O~II] line emissions \citep{Fesen06}, exhibit the O/S sensitive
F775W/F850LP ratio several times greater than even strong [O~II]
$\lambda\lambda$7319,7330 emission main shell knots.

Specifically, we chose a flux ratio for F775W/(F625W + F850LP) $\geq$ 1.0 to
separate out the [O~II] strong FMKs from the [S~II] strong FMKs; that is, those
knots with stronger [O~II] $\lambda\lambda$7319,7330 emission detected via the
F775W filter than the combined strength of [N~II], [S~III] and [S~II] emissions
detected in the F625W and F850LP filters.  

Similarly, knots with strong [N~II]
emissions were selected via F625W/(F775W + F850LP) $\geq$ 1.0, thereby
selecting those knots where the combined flux of [O~I] 
$\lambda\lambda$6300,6364, [S~II] $\lambda\lambda$6716,6731, and [N~II]
$\lambda\lambda$6548,6583 emissions was greater than the sum of F775W flux, due
mostly to [O~II], and F850LP flux sensitive to the near-infrared [S~III] and
[S~II] emissions. 
Since [O~I] $\lambda\lambda$6300,6364 flux rarely, if ever, exceeds the [O~II]
$\lambda\lambda$7319,7330 flux, and the observed [S~II]
$\lambda\lambda$6716,6731 emission is unlikely to ever exceed the combined flux
of [S~III] $\lambda\lambda$9069,9531 plus [S~II] $\lambda\lambda$10287--10370
line emissions \citep{Hur96,Winkler91}, then any knot for which the
F625W/(F775W + F850LP) $\geq$ 1.0 requires the presence of significant [N~II]
$\lambda\lambda$6548,6583 emission.

\section{Results}

Examination of March and December 2004 ACS/WFC images revealed a total of 1825
high proper motion ($\mu \geq 0\farcs35 - 0\farcs90$ yr$^{-1}$) ejecta knots
around the Cas~A remnant \citep{HF06}.  This is a much larger population of
high-velocity ejecta knots than previously identified or suspected from
ground-based images \citep{Kvdb76,Fes96,Fesen01}.  A total of $444$ strong
[N~II] emission, $192$ strong [O~II] emission, and $1189$ FMK-like knots were
identified.  Although nearly half of the 1825 cataloged outer knots were found
in the NE jet, high-velocity outlying ejecta were identified in many other
regions around the remnant.

Outlying optical ejecta knots range from 105$''$ to 300$''$ in radial distance
from the center of expansion, placing them in projection close to or outside
the remnant's $\simeq$ 6000 km s$^{-1}$ forward shock front as determined by
the remnant's outermost X-ray emission \citep{Gotthelf01,Delaney03}.  The
location and distribution of these knots with respect to the remnant's outer
X-ray emission filaments associated with the forward shock front can be seen in
Figure 1.  Here we show the locations of all 1825 cataloged outer knots
projected onto the 1 Msec {\sl Chandra} ACIS image (epoch 2004.3;
\citealt{Hwang04}).  Knot positions are marked with open circles color coded
either red, green, or blue to indicate those knots with spectra dominated by
strong [N~II] $\lambda\lambda$6548,6583, [O~II] $\lambda\lambda$7319,7330, or
[S~II] $\lambda\lambda$6716,6731 line emissions, respectively. 

The transverse expansion velocities of Cas~A's outermost ejecta (Fig.\ 1) show
a strongly aspherical structure due principally to the NE and SW jets which
appear as distinct and roughly opposing features; NE jet: P.A. = $45{\deg} -
70{\deg}$, SW jet: P.A. = $230{\deg} - 270{\deg}$. The jets have similar
maximum radial distances and proper motion derived expansion velocities;
namely $r = 290''- 300''$  and $v = 14000$ km s$^{-1}$.  They also contain the
remnant's highest velocity [S~II] emitting ejecta (the blue open circles in Fig.\ 1).
This is in contrast to other regions where the nitrogen-rich ejecta (open red
circles) represent the remnant's highest velocity debris ($10000 - 11000$ km
s$^{-1}$) followed by the O-rich ejecta, and then the S-rich knots
\citep{Fesen06}.

Examination of the {\sl HST} images also revealed a lack of outlying ejecta
knots along the remnant's northern and southern regions. Not a single ejecta
knot could be found along a 55$\deg$ wide region in position angle along the
remnant's southern limb (i.e., PA = $145\deg - 200\deg$) 
and in a narrower 15$\deg$ position angle zone along
the north (PA = $335\deg - 350\deg$). 
The northern gap would actually be larger if one only considered
knots with radial distances greater than $180''$ in which case the northern gap
size grows to $\sim35\deg$ in position angle (PA = $327\deg - 3\deg$).  We
note that, unlike for the northern limb region, ACS images did not
completely cover the whole southern limb region imaging only out to a distance
of $180''$ due south of the center of expansion.  
However, inspection of ground-based images covering regions farther to
the south ($\sim200''$) revealed no bright farther outlying knots.
Thus, despite  missing ACS/WFC coverage along the south,
the lack of any knots at smaller distances like
those seen along the east and west edges of the southern gap (r = $150'' -
185''$) indicate a strong likelihood that there are simply no bright, high-velocity
ejecta knots in either of the NNW and SSE directions. 

This NE--SW jet/counterjet and NNW--SSE gap asymmetry in the distribution of the
remnant's outer optical ejecta knots can be clearly seen in the proper motion
plots of Figure 2.  In the top panel, we plot the extrapolated 320 yr proper
motion paths for the 1825 identified outer knots based on measured March to
December 2004 positions.  The central white circle has a radius of $5''$
centered on the remnant's estimated center of expansion (COE; $\alpha$(J2000) =
$23^{\rm h} 23^{\rm m} 27\fs77 \pm 0\fs05$, $\delta$(J2000) = $58\deg 48'
49\farcs4 \pm 0\farcs4$; \citealt{Thor01}).  Although errors in derived knot
proper motions lead to some knot trajectories missing the expansion center by
substantial distances filling in somewhat the northern and southern gaps, a
strongly aspherical distribution of ejecta jets and gaps is readily
apparent.

When the proper motion extrapolations are replaced with predicted knot proper
motions based solely on March 2004 knot positions and the COE, a structure of
opposing jet features and north and south gaps appears even more striking
(Fig.\ 2, bottom panel). The circle in the figure marks a radial distance of
$200''$ from the COE corresponding to a proper motion of $0\farcs625$ yr$^{-1}$
for an age of 320 years and an implied $\simeq$10,000 km s$^{-1}$ transverse
velocity assuming a remnant distance of 3.4 kpc. As can be seen from the
figure, this circle encompasses nearly all the remnant's N-rich outer ejecta
away from the jet regions. 

Peak NE and SW jet optical knot transverse velocities lie at PA = $60\deg -
61\deg$ and at PA = $238\deg - 243\deg$, respectively, i.e., $\simeq$180$\deg$
apart. These position angles are close to those seen also in the X-rays and
infrared for the NE and SW jets. That is, from the 1 Ms {\sl Chandra}
\citep{Hwang04} and 24 micron {\sl Spitzer} images \citep{Krause05}, we find PA
= 63$\deg$--65$\deg$ for the center of the NE jet and PA = 246$\deg$--248$\deg$
for the center of the SW jet, some $183\deg$ apart.  While the significance of
the northern and southern gaps in the distribution of the outer knots is
uncertain (see below), they are situated roughly orthogonal to this NE--SW jet
alignment line.  The middles of the southern and northern outer knot gaps lie
at PA = $170\deg$ and $342\deg$ respectively, some 107$\deg$ and 99$\deg$ off
from a 63$\deg$--243$\deg$ alignment line.   

Lastly, extrapolated late 17th century knot positions relative to the remnant's
estimated COE indicates some evidence for non-uniform deceleration around the
remnant.  Figure 3 (left panel) shows the estimated positions for all 1825
outer knots for the \citet{Thor01} estimated convergent year 1671.  The
estimated COE is marked by the circle (radius = $5''$).  Here, one sees that
the scatter of points for the outer knots is significantly away from the
\citet{Thor01} center, centered instead $3\farcs4$ to the east.  (Note: A check
using the same 17 outer knots employed by \citet{Thor01} using the new ACS
images together with archival images agree with their derived COE.) One
possible cause for this shift eastward is a greater deceleration of knots along
different regions around the remnant, particularly along the eastern limb,
giving rise to broader and slower average expansion velocities for a given
radial distance and look-back time.  A different shift off from the COE is seen
when plotting just northern and southern knots (Fig.\ 3; right panel). For
416 north and south knots, the shift from the COE shows an even greater
displacement, namely $5\farcs2$ to the southeast.

\section{Discussion}

Although optical debris may constitute only a small fraction of the total
ejected mass, Cas~A's fastest moving material is perhaps best studied
optically.  For example, optical emission from the NE and SW jets can be traced
about $80''$ farther out than in X-rays or infrared \citep{Hwang04,Krause05}
and only a handful of outer optical ejecta knots around the rest of the remnant
are detected in even the deepest radio or X-ray images.

Fragmentation of SN ejecta into dense clumps may come about by Rayleigh-Taylor
instabilities brought on by the deceleration of the ejecta by the ambient
medium \citep{Gull75,Jones94,Jun96}, passage of the reverse shock \citep{Herant94}, or
inside the progenitor during the SN explosion \citep{Chevalier75,Chevalier78}.
Optical emission will arise when $\sim$100 km s$^{-1}$ internal shocks are
formed in the clumps driven by the high stagnation pressure behind the clump's
bow shock.

\citet{Hamilton85} studied similarity solutions for SN blast waves driven by
clumpy ejecta and found that such clumps will eventually move ahead of a
remnant's forward shock front. He concluded that this process may be a generic
feature of SNRs with clumped ejecta.  The presence of numerous dense and
relatively cool optically emitting, high-velocity ejecta knots out ahead of
Cas~A's forward shock front supports Hamilton's basic results. Namely, we find
a significant number of optical outer ejecta knots lying (in projection)
outside the faint X-ray emitting filaments marking the current location of the
remnant's forward blast wave \citep{Gotthelf01,Delaney03}.

\subsection{Asymmetrical Expansion}

Asymmetries in Cas~A's outermost debris may offer clues as to the nature of the
SN explosion engine. What one observes at a particular epoch, however, may not
necessarily reflect the SN's true explosion dynamics.  
Ejecta velocities will be modified (to varying degrees) by
passage through the ambient medium as ejecta knots are only made optically
visible due to this interaction, and more knots will be visible in regions
where there is more circumstellar material.

The detected distribution can also change with time since some 10--20\% of
outer optical knots vary significantly in brightness over just a few years due
presumably to variable interaction with a clumpy CSM \citep{HF06}.  In
addition, for a knot to be visible it must be sufficiently large and dense to
generate detectable forbidden line emission.  It must also survive disruption
via mass ablation from Kelvin-Helmhotz instabilities during passage through the
reverse shock, forward shock, and the surrounding ambient medium.
Consequently, what one sees can be a distorted and biased view of the true
distribution of a remnant's outer debris.  

Nonetheless, even with all these caveats, the asymmetry and non-uniformity of
the remnant's outermost optical ejecta as seen in the plots of Figure 2 is
striking.  The NE and SW jets extend out to nearly $300''$, more than twice the
radius of the remnant's main emission shell, and lie on virtually opposite
sides of the remnant's center of expansion.  The jets also appear to lie
roughly in the plane of the sky to within $\sim30\deg$
\citep{min68,Fes96,Fesen01}.  This combination of an opposing, sulfur-rich
jet/counterjet structure with similar maximum expansion velocities plus gaps of
high-velocity ejecta aligned roughly perpendicular is suggestive of an
aspherical, bipolar expansion.

From 2-D models, \citet{Blondin96} calculated that a progenitor's axisymmetric
wind structure, where the highest density lay in the equatorial plane, could
generate bipolar protrusions of SN ejecta extending $2 - 4$ times the radius of
the main shell.  The relative scale of such expansion asymmetry is not unlike
that seen in the Cas~A jets and only a mild asymmetry ($\rho_{e}$/$\rho_{p}$
$\sim$ 2) is needed to form obvious protrusions. While significant heavy
element material can end up at much larger radial distances than elsewhere, the
original progenitor chemical layering should largely persist within the
protrusion. That is, the highest velocity material should mainly exhibit the
chemical abundances of the progenitor's outermost layers, namely N, He, and O.
However, just the opposite is observed. The farther out one goes in either of
Cas~A's jets, the weaker oxygen emission lines become relative to those of
sulfur \citep{vdb71,Fes96,Fesen01}.  

The fact that one sees largely undecelerated ejecta throughout the
remnant with $ v \propto r$ (see \citealt{Fesen06}) indicates that the observed
compositional structure is set early in the explosion. In addition, outside of
the jet regions, the S-rich material has a much more limited velocity range.
All this suggests that the S-rich NE jet, rather than reflecting some sort of
turbulent mixing region or a rapid expansion into surrounding ISM/CSM due to a
lower surrounding density, more likely represents a true stream of high-speed
ejecta formed when underlying material was ejected up through the star's outer
layers.

Other observational evidence in favor of an ejection of underlying material
comes from the location of the so-called `mixed ejecta knots' and from energy
estimates in the jet region.  Mixed ejecta knots show both nitrogen and sulfur
overabundances, which could have resulted from turbulent mixing of H and N-rich
layers with underlying S and O-rich layers, are only seen in the jet regions
\citep{FB91,Fesen01}. The ACS/WFC images show that many of the mixed knots are
not the result of line-of-sight superposition of knots with different chemical
make-up but appear to be ejecta with mixed chemical properties \citep{HF06}.
In addition, from an analysis of {\sl Chandra} X-ray data, \citet{Laming03}
found a shallower outer envelope near the base of the NE jet which they
interpreted as indicating more of the initial explosion's energy being directed
in this polar direction as opposed to equatorial.  On the other hand, however,
while the prominent shell rupture near the base of the NE jet seen in X-rays
and radio images might be seen as supporting evidence for this picture
\citep{Fes96}, there is no obvious main shell rupture near the SW counterjet.  

\subsection{Jet/Counterjet: MHD or Neutrino Driven?}

Although our {\sl HST} survey of the highest-velocity ejecta in Cas~A makes a
compelling case for a high-velocity bipolar expansion, what the jets and
gaps are telling us about the underlying core-collapse mechanism is 
unclear.  

Currently, there is no consensus as to the details of the explosion process for
core-collapse SNe except that models with completely spherical neutrino heating
mechanisms fail to yield successful explosions
\citep{Rampp00,Lie01,Lie05,Thompson03}.  This has led to considerable effort
to understanding the effects of rotation on neutrino heating and the
importance of magnetic fields in the proto-neutron star.

Around the rotation axis of a collapsing star, accretion flow will be
non-spherical and accelerated thereby lowering the needed critical neutrino
luminosity \citep{Yamasaki05}.  The degree of neutrino re-vitalization of the
shock front created by the core-bounce may also be affected by convection in
the proto-neutron star through the mechanical outward transfer of neutrinos.
Investigations into the effects of rotation on the anisotropic neutrino
emission from the proto-neutron star indicate a weakening of the core bounce
that seeds the neutrino-drive convection, with angular momentum tending to
stabilize the core, constraining convection to the polar regions
\citep{Fryer00,Burrows04}.  Models of anisotropic neutrino radiation indicate
more powerful explosions are generated, which then lead to a prolate expansion
geometry \citep{Shimizu01,Mad03,Mad04,Wilson05}.  \citet{Walder05} found
the bipolar expansion was not strongly collimated ($\sim30\deg - 60\deg$)
unless the rotation rate was large.

For Cas~A, \citet{Burrows04,Burrows05} favored a rotation enhanced, neutrino-driven model
in which the X-ray observed Fe-rich SE and NW regions, and not the NE
and SW jet/counterjet, mark the progenitor's rotation axis.  Instead of driving
the explosion, the NE--SW jets would have formed following the neutrino-driven
main explosion via an under-energetic jet-like ejection created by an MHD jet or
proto-neutron star wind (possibly associated with accretion by the NS;
\citealt{Janka05}) emerging into an already expanding debris field. But this
model requires a nearly 90$\deg$ post-explosion precession of the NS's rotation
axis from NW--SE to NE--SW \citep{Burrows04,Burrows05}, the cause of which is
left unexplained. 

On the other hand, models of magnetorotationally induced jets capable of
generating SN explosions have also been invoked to explain
the Cas~A jets \citep{Khok99,Wheeler00,Wheeler02,Akiyama03,Tak04}. In this
view, rotation leads to magnetic field amplification thereby generating
non-relativistic axial jets of MHD energy  $\sim10^{51-52}$ erg which then
initiate a bipolar supernova explosion. This is somewhat analogous to the
narrow relativistic jets proposed in the collapsar model as the central engine
for GRBs.  \citet{Kifonidis03} questioned the accuracy of anisotropic jet
explosion simulations and \citet{Janka05} argue that if the Cas~A jets were
driven by outflowing core material they should be Fe-rich instead of the S and
Si rich material observed. 

If Cas~A's NE and SW jets were driven by Ni-rich material, the energy
deposited by radioactive $^{56}$Ni decay might have created hot, low density
Ni-rich bubbles. This could make Fe-rich material in the jets today too
cool for detection in X-rays while also being too diffuse and low density to 
detect as optically bright knots.  
Furthermore, while both the NE and SW jets can
be traced farthest out in the optical, optical studies have never detected
appreciable Fe-rich material anywhere in Cas~A. Optical Fe line emissions are
weak and hard to detect in even the brightest main shell knots and there is no
optical Fe-rich material seen corresponding to the Fe-rich SE and N regions
observed in the X-rays \citep{Winkler91,Reed95,Hur96}.

Morphologically, both MHD jet and neutrino-driven expansion models produce
aspherical `jets' with axial expansion ratios around 2 like those seen for
Cas~A \citep{Khok99,Kotake05}.  Although much of the NE jet's central and farthest
extending line of optical filaments lie virtually in the plane of the sky
\citep{min68,Fes96}, knots lying at a projected radial distance of 150--170 arcsec
out from the COE show radial velocities from $-3000$ to $+5000$ km s$^{-1}$,
indicating an expansion cone about $25\deg$ wide \citep{Fes96}. This means that
the NE jet is more like a fan of several streams of ejecta knots (rather than a
single narrow jet), about as deep as it appears wide, i.e., about $\sim
25\deg - 35\deg$.  This is consistent with their appearance on X-ray and
infrared images which show both jets as two or three ejecta streams rather than
one single narrowly focused line of emission. For example, there are three main
optical lines of knots in the NE jet which correlate with X-ray and IR emission
fingers. Although the SW jet is optically much less well-defined, the locations
of the outermost [S~II] emitting ejecta correlate reasonably well with the X-ray and IR
emission `fingers'.

Beside the jets, a possible additional clue as to the nature of the central
core-collapse SN engine may be the absence of high-speed ejecta along the north
and south limbs.  Not all proposed core-collapse models show a pronounced
decrease in outermost expansion velocity suggested by the nearly opposing
ejecta gaps revealed in the {\sl HST} data (Fig.\ 2).  However,  similar gaps
in the distribution of slower moving ejecta are not found in the remnant's main
emission ring of reverse shocked debris based on radio, X-ray, and optical
maps. While the remnant's forward shock shows no decrease in these
directions as one might expect if the expansion velocity was significantly
lower in these northern and southern gap regions \citep{Gotthelf01},
a gap in the forward shock front would be rapidly filled by the blast wave 
advancing in from the sides.  Thus,
while the north and south ejecta gaps are interesting especially given their
positions relative to the jets, their meaning is presently unclear.
Nevertheless, if these gaps are truly found to be areas devoid of high-speed
ejecta seen elsewhere around the remnant and orthogonal to the NE/SW jets they
may provide some insight for testing core-collapse models.

\subsection{Motion of Central X-ray Point Source}

First-light images of Cas A taken by the {\sl Chandra} X-ray Observatory
revealed a central X-ray point source (XPS) in the remnant \citep{Tananbaum99}.
Although its nature is uncertain, this object is likely to be a neutron star
but not a pulsar due to a lack of radio pulsations, no detected X-ray or radio
plerion, and an X-ray spectrum too steep for an ordinary pulsar
\citep{Pavlov00}.  It has been suggested that it may be a younger and less
luminous example of a subclass of neutron stars known as anomalous X-ray
pulsars (AXPs) and soft gamma-ray repeaters (SGRs)
\citep{Chak01,Mere02,Pavlov02,Roth03,Pavlov04,FPS06}.  X-ray emission bursts
from AXPs and SGRs together with their spin-down rates have been explained by a
magnetar model in which a neutron star has a much higher surface magnetic field
of $10^{14}-10^{15}$ G than ordinary pulsars \citep{DT92,Thompson96}.  Except
for a lack of pulsations, the general properties of the Cas A XPS and other
X-ray emitting but radio quiet compact central objects in fairly young SNRs are
not all that dissimilar from AXPs and SGRs \citep{Pavlov02,FPS06}.

Core-collapse asymmetries in the SN explosion might help explain the inferred
high space velocities ($200 - 500$ km s$^{-1}$ or more) for neutron stars and
radio pulsars \citep{LL94,CC98,B03}. Although 3D core-collapse models suggest
neutrino asymmetries as well as disruption of binaries by symmetric explosions
may be insufficient to generate the wide range of observed `kick' pulsar
velocities \citep{CC98,Fryer04}, MHD driven explosions creating unbalanced jets
just might \citep{Khok99}.

However, the motion of the remnant's XPS is far from being in alignment with
Cas~A's NE--SW jets.  Using the initial {\sl Chandra} derived position along
with subsequent positional measurements using archival {\sl ROSAT} and {\sl
Einstein} image data \citep{Aschen99,Pavlov99}, \citet{Thor01} estimated a $6
\farcs 6$ displacement of the XPS from their derived center of expansion.
Assuming a common origin for the XPS and the expanding ejecta, they estimated a
transverse velocity of $\approx$330 km s$^{-1}$ for a distance of 3.4 kpc.
Adopting an updated position of the XPS (2004 epoch; \citealt{FPS06}), the 
displacement of the XPS from the \citet{Thor01} remnant expansion
center is $7 \farcs 0 \pm 0 \farcs 8$ with an implied motion in a
southeasterly direction (position angle = $ 169{\deg} \pm 8.4 {\deg}$; see
Fig.\ 4, left panel). Assuming a current remnant age $\simeq$ 325 yr (see
discussion in $\S$5.4), we find a slightly higher
transverse velocity of around 350 km s$^{-1}$ for a distance of 3.4 kpc.

The apparent southernly direction of motion for the XPS is roughly orthogonal
to the jet-counterjet alignment line, making an asymmetric jet-induced `kick'
explanation problematic.  Interestingly however, the projected motion of the
XPS is toward the middle of the broad southern gap in the distribution of the
outer ejecta knots (Fig.\ 4; right panel).  If the distribution of the
optically emitting outer ejecta knots is giving us a true picture of the
asymmetry in the Cas~A supernova expansion, then the neutron star's preference
to move in a direction lacking in high velocity material may indicate a natal
kick aligned with the progenitor's slowest velocity expansion, possibly
the progenitor's equatorial region. 

\citet{DT92} and \citet{Arras_n_Lai99} have proposed that magnetars could
receive natal kicks in part due to their intense magnetic fields which could
lead to anisotropic neutrino emissions. However, unlike the case seen here for
Cas A, these kicks would be in the direction of the rotation axis -- which the
NE--SW jets would seem to mark. While some misalignments between the
progenitor's spin axis and a neutron star's velocity vector have been reported
\citep{HB99}, including the high velocity pulsar B1508+55 (V$_{\rm trans}$
$\simeq$ 1100 km s$^{-1}$; \citealt{Chat05}), there is not strong observational
evidence for general misalignments \citep{Desh99}.  Moreover, although some binary
disruption type models have been proposed to explain natal kicks perpendicular
to the spin axis (e.g., \citealt{Wex00,CW02}), no clear picture has emerged on
how such a misalignment would be produced or even whether such models apply to
the case of Cas~A. 
                                                                                                                                   
\subsection{Dynamical Age Estimates for Cas A}

Although only nine months separated the two sets of ACS/WFC images, the large
number of outer knots identified (1825) permitted us to estimate the dynamical
age of the Cas~A SNR.  Assuming the COE derived by \citet{Thor01} and no knot
deceleration since the time of the explosion, we show in Figure 5 (left panel)
the estimated arrival date of the 1825 cataloged outlying ejecta knots to
within the minimum least squared distance to the COE plotted versus cataloged
knot identification number \citep{HF06}.  In this figure, symbol size is
inversely proportional to estimated proper motion uncertainty.  Knot catalog
IDs are in order of increasing position angle with the NE jet knots (ID numbers
$\sim$ 100 -- 950) making up a substantial fraction ($\sim$45\%) of the 1825
cataloged knots.

The dispersion in estimated knot convergent dates is not uniform as a function
of position angle. For example, for the middle section of the NE jet where many
of the fastest ejecta are found (i.e., Knot IDs $350 - 550$), a decrease in the
range of estimated knot arrival dates can be seen in Figure 5 (left
panel).  This decrease reflects more accurate proper motion values due mainly
to the larger radial distances for jet knots from the COE and consequently
larger proper motion values leading to smaller percentage measurement errors.

\subsubsection{Age Estimates Assuming No Knot Deceleration}

The average arrival date for the 1825 outer knots with undecelerated
extrapolated arrival dates between 1580 and 1750 is 1662$\pm$27 yr.  This is
consistent with that estimated by \citet{Thor01} who found an undecelerated
convergent date of 1671.3$\pm$0.9 based on a sample of 17 especially long-lasting knots
for which archival imaging data were available covering a time span of up to 50
years.  Because the catalog of 1825 outer knots covered a wide range of sizes
and brightnesses leading to a range of proper motion measurement errors, we
examined a much smaller, hand-selected sample of 72 knots which are: i)
relatively bright, and ii) compact in size or unresolved in the ACS/WFC images.
This sample included the 17 outer knots used by \citet{Thor01} for their
estimated convergent date.  The results for this smaller sample, shown in
Figure 5 (right panel), indicate a convergent date $1671.8 \pm 17.9$,
in excellent agreement with the Thorstensen et al.\ 1671 date, shown here
as a vertical dashed line.

\subsubsection{Knot Deceleration}

While ejecta knots must undergo shock heating and hence some deceleration in
order to be optically visible, the least decelerated ones offer stronger 
upper limits to the remnant's age.  As seen in Figure 5, the region near the top
of the left-hand plot, i.e., knot IDs from $1690 - 1815$ corresponding to
position angles $275{\deg} - 315{\deg}$ respectively, show displacements toward
a convergent date later than 1671, namely $1680.5 \pm18.7$. These 126 knots are
located along the remnant's northwest limb and can be seen in the bottom panel
of Figure 2 as cluster of NW knots.

If we knew the degree of deceleration these northwestern limb
knots might have experienced over the $\sim$300 yr
age of the remnant, it would give us a better estimate of the remnant's age
and therefore the actual Cas A SN explosion date. For example, if knot
decelerations were significant then one might be able to
rule out on dynamical grounds a proposed but controversial sighting of
the Cas A SN by Rev.\ John Flamsteed in August 1680 \citep{Ashworth80,SG03}.  

A knot will undergo deceleration both from the direct interaction with local
gas as well as from the internal shock driven into the knot that gives rise to
the optical emission observed \citep{Jones94}.  If treated as a dense
undistorted clump, a knot's deceleration due to drag from its interaction with
the ambient medium depends on knot velocity and mass, the density of the local
medium, and the cross-sectional area of the knot's bow shock which for hypersonic
conditions is approximately equal to the knot itself \citep{Chevalier75,Hamilton85,Jones94}.

The timescale for knot deceleration (drag) is given by
$\tau_{drag}$ $\sim$ $\chi R_{k}$/$v_{k}$ where $\chi$ is the density contrast between
the knot and the ambient medium (i.e.,  $\rho_{k}$/$\rho_{a}$), $R_{k}$ is the
knot's radius, and $v_{k}$ is the knot's velocity \citep{Jones94}.
Based on our ACS imaging data, typical outer
knots have 
velocities of 10,000 km s$^{-1}$ and sizes
R $\simeq 0\farcs1 $ corresponding to 0.002 pc at 3.4 kpc.
High-velocity outlying [S~II] emitting knot electron densities lie between 2000 -- 16,000 cm$^{-3}$
with typical values between 4000 -- 10,000 cm$^{-3}$
\citep{Fes96,Fesen01}. The ambient density around Cas~A is not well determined but is estimated
to range between 0.4 -- 3.7 cm$^{-3}$ \citep{Braun87}. 

Choosing a $\chi$ = 10$^{4}$ and an outer ejecta knot velocity of 10,000 km
s$^{-1}$ leads to $\tau_{drag}$ $\sim$ 2000 yr, suggesting that outer knot
deceleration due to drag may be fairly small at present. This conclusion is
consistent with a lack of detectable knot deceleration over the last 50 yr and
a velocity change of $<$ 5\% over 300 yr \citep{Thor01}. While model
simulations  suggest that shocked clumps become flattened (i.e., laterally
spread) which would increase their cross section and hence enhance the
deceleration, radiative cooling might partially counteract this effect leading
to clump breakup into a cluster of smaller, denser knots.

Cloud-ISM interaction models suggest knot disruption might also occur due
to Rayleigh-Taylor and Kelvin-Helmholtz instabilities, again resulting in the
generation of smaller, dense knot fragments
\citep{Klein94,Jones94,CidFern96}.  The timescale for initial knot breakup
under these conditions, $\tau_{b}$, is uncertain but is likely to be a few
`cloud crushing times' \citep{Klein90,Jones94} or $\tau_{b}$ $\sim$ 4
$\chi^{1/2} R_{k}$/$v_{k}$. For the knot numbers assumed above, the disruption
timescale is $\sim$ 50 yr, meaning the remnant's highest-velocity ejecta clumps
might well have already undergone breakup into denser knots. Thus, both radiative
cooling effects and dynamical instabilities might help explain the observed
small-scale clustering of some of the remnant's outer ejecta knots (see \citealt{HF06}).

Knot deceleration from internal shock passage will cause the largest error in
convergent date if it has occurred recently, and given the $\sim20 - 30$ year
lifetimes of most optical knots in Cas A \citep{Kvdb76}, this is probably the case.  The
effective shock deceleration is equal to the internal knot shock velocity for
post-shock emission \citep{Dopita95}.  The shock speed must be at least 30 $\rm
km~s^{-1}$ to produce significant [N~II], [S~II] and [O~II] emissions (e.g..
\citealt{Hartigan94,Blair00}).  Based on the Blair et al. (2000) shock models, shock
speeds above about 500 $\rm km~s^{-1}$ will ionize the oxygen to the
helium-like ionization state and higher, reducing the cooling rate by an order
of magnitude.  If the outer portions of the SNR are roughly in pressure
equilibrium, the density is proportional to v$^{-2}$, further reducing the
cooling rate behind fast shocks.  Consequently, shocks above about 500 $\rm
km~s^{-1}$ will have cooling times longer than the age of Cas A, and will
appear as X-ray rather than optical emission features.   Thus the observed
knots have been decelerated by 30 to 500 $\rm km~s^{-1}$, or 0.3 to 5\% of
their $\sim$10,000 km s$^{-1}$ apparent speeds.  This corresponds to an error
in convergence time of 0.003 to 0.05 times 320 yr, or 1 to 15 yr, well within
the 19 yr uncertainty in the convergence time of the least decelerated
northern limb knots.  

We conclude that our convergence times estimates are not likely to be
significantly affected by knot deceleration. Therefore, while the seemingly less
decelerated knots located along the remnant's northern limb suggest an
explosion date somewhat later than 1670, overall the measurements are still consistent
with a possible sighting of the Cas A SN in 1680. Follow-up ACS imaging
obtain in a few years may be able to settle this issue more definitely through
a firmer estimate of the date for the Cas A supernova outburst.

\acknowledgments

This work was supported by NASA through grants GO-8281, GO-9238, GO-9890, and
GO-10286 to RAF and JM from the Space Telescope Science Institute, which is
operated by the Association of Universities for Research in Astronomy.
RAC is supported by NSF grant AST-0307366 and CLG is supported through 
UK PPARC grant PPA/G/S/2003/00040.

\clearpage

\newpage

%
%

\begin{figure}
\epsscale{0.85}
\plotone{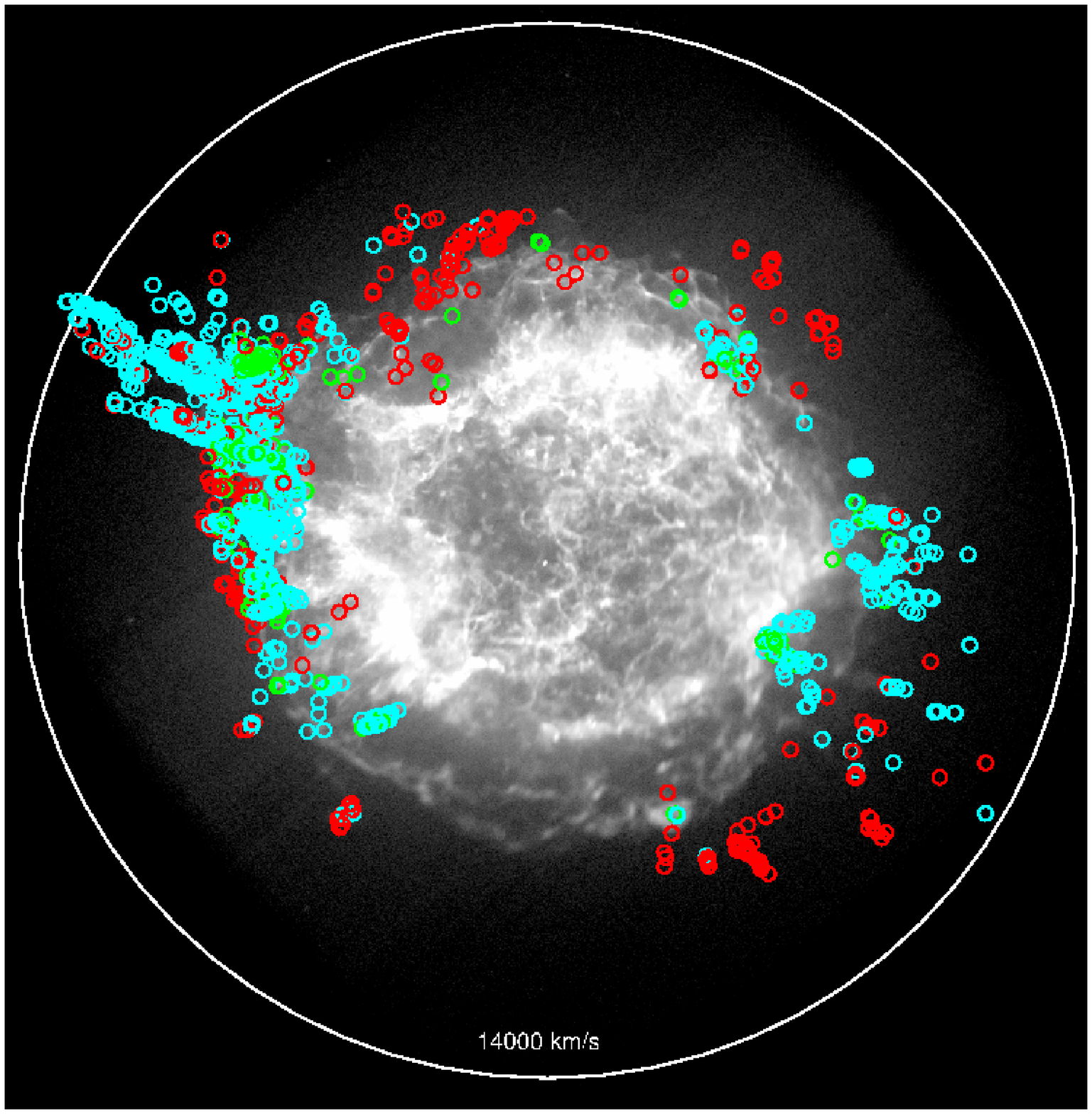}
\caption{The 1~Msec {\sl Chandra} image of Cas A with the locations of the 1825 identified outer
ejecta knots \citep{HF06} marked color coded by their emission properties. Red open circles indicate
knots with strong [N~II] line emission, green open circles knots strong [O~II] emission,
and blue open circles strong [S~II] FMK-like outlying knots.}
\label{fig:X-ray_with_knots}
\end{figure}

\begin{figure}
\epsscale{0.65}
\plotone{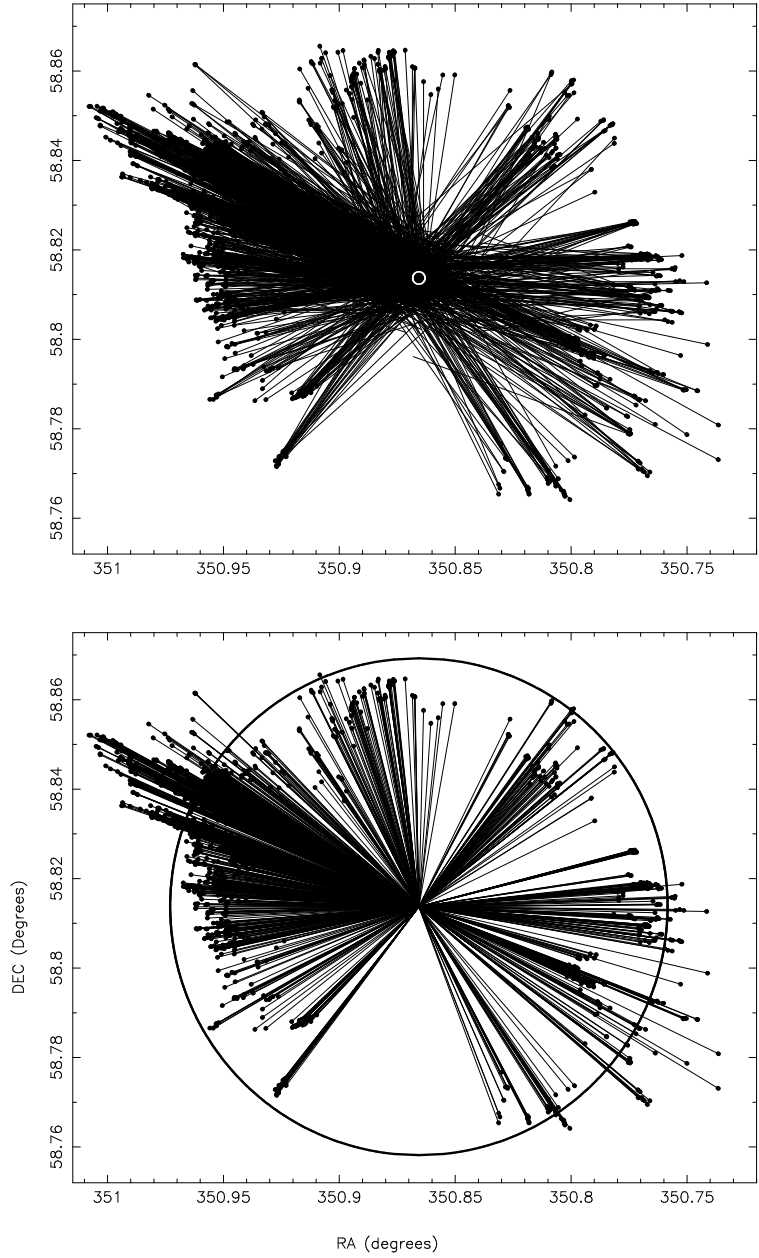}
\caption{{\it Top:} Plot of extrapolated 320 yr proper motions 
for the 1825 identified outer knots 
based on actual proper motions measured using the March and December 2004 ACS/WFC data. 
Central white circle has a radius of $5''$ and marks the remnant's estimated center of
expansion \citep{Thor01}.  
{\it Bottom:} Plot of 1825 outer knot positions and their expected motions away from the remnant's 
known center of expansion
revealing a `bow-tie' asymmetric structure. Circle represents the radial 
distance of $200''$ corresponding to a measured
proper motion of $0\farcs65$ yr$^{-1}$ and thus an implied 10,000 km s$^{-1}$ 
transverse velocity at the assumed remnant distance of 3.4 kpc.
}
\label{fig:kaboom}
\end{figure}

\begin{figure}
\epsscale{0.90}
\plottwo{f3a.ps}{f3b.ps}
\caption{
{\it Left Panel:} Plot of extrapolated 1670 epoch positions for 1825
                  outer ejecta knots relative to the \citet{Thor01} estimated center of
                  expansion (COE) marked by the grey (red) circle (radius = $5''$)
                  centered on $\alpha$(J2000) = $23^{\rm h} 23^{\rm m} 23^{\rm s}.77$, $\delta$ = $50{\deg} 48' 49\farcs4$.
                  The scatter of points is centered $3\farcs4$ from the COE.
{\it Right Panel:} Same as for the left panel except that knots located along the east and west limbs of
the remnant have been removed, leaving 416 north and south knots. The scatter of these knots is 
centered $5\farcs2$ from the COE.}
\label{fig:knot_coe}
\end{figure}

\begin{figure}
\epsscale{0.90}
\plottwo{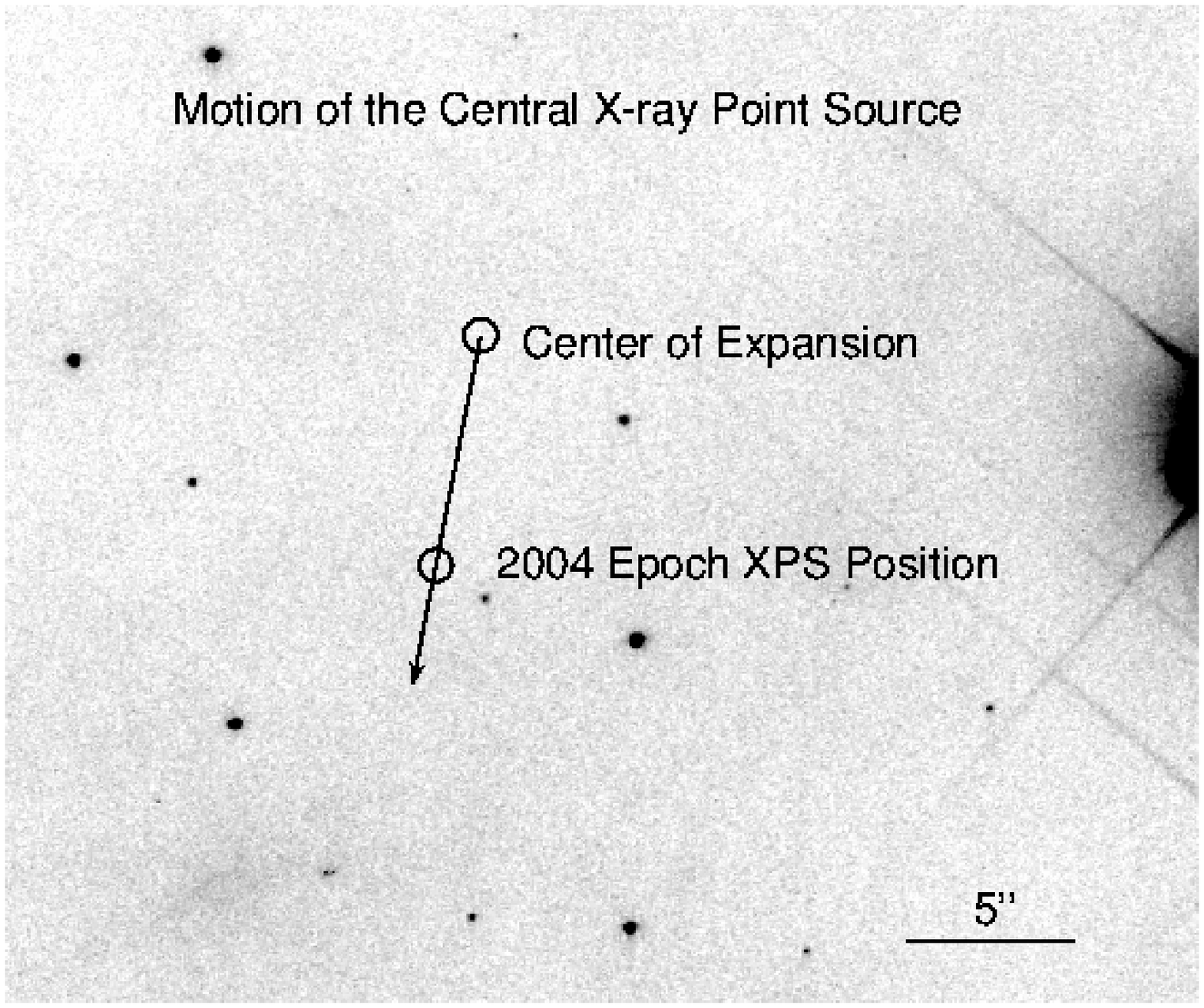}{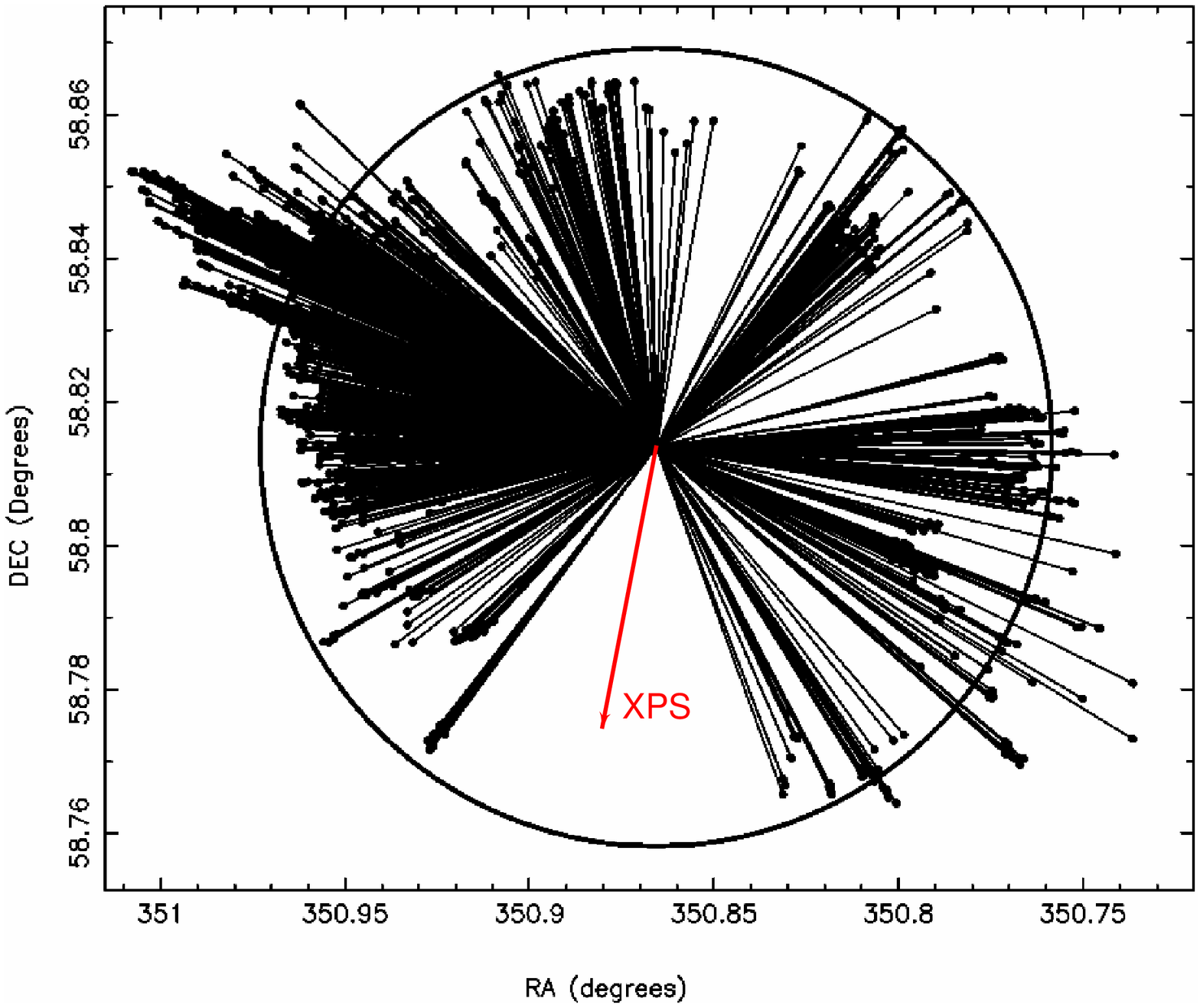}
\caption{
{\it Left Panel:} A 2001 STIS {\sl HST} image of the central region of Cas~A near the X-ray Point Source (XPS)
with the \cite{Thor01} expansion center marked ($\alpha [J2000] = 23^{\rm h} 23^{\rm m} 27 \fs 77\pm 0\fs 05$,
$\delta [J2000] = 58^{\rm o} 48' 49 \farcs 4 \pm 0\farcs 4$) along with the XPS's current position
 ($\alpha [J2000] = 23^{\rm h} 23^{\rm m}27 \fs 943\pm 0\fs 05$,
$\delta [J2000] = 58^{\rm o} 48' 42 \farcs 51 \pm 0\farcs 4$)
as derived from {\sl Chandra} image data (see \citealt{FPS06}). The circles marking these positions
are 1$''$ in diameter. The separation between the remnant's estimated expansion
center and the XPS' current position is $7\farcs0 \pm0\farcs8$ with an implied motion in a southeasternly direction
(position angle = $169{\deg} \pm 8.4{\deg}$).
{\it Right Panel:} Same plot as shown in the bottom panel of Figure 2 but now showing the apparent motion of Cas~A's XPS in the
direction of the southern gap of high-velocity, outer ejecta knots.
}
\label{fig:xps_motion}
\end{figure}
                                                                                                                                                      
\begin{figure}
\epsscale{0.90}
\plottwo{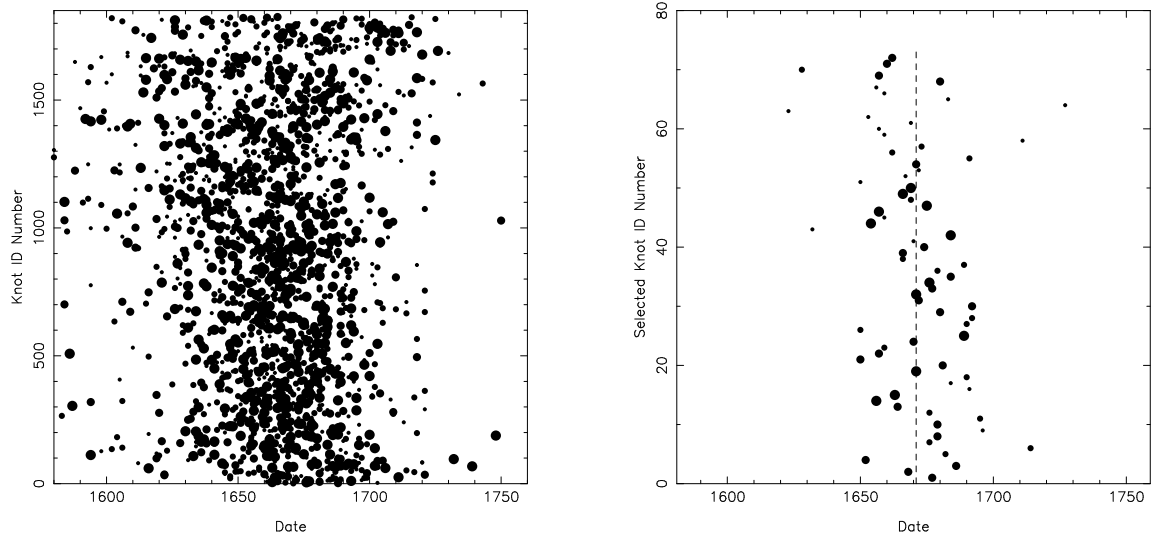}{f5b.ps}
\caption{
{\it Left Panel:} Plot of catalog knot identification number (ordered in position angle) with the date of closest approach
to remnant's estimated center of expansion (COE). 
Symbol size is inversely proportional to estimated proper motion uncertainty.
{\it Right Panel:} Same as for the left panel but now showing date of closest approach to the COE 
for 72 selected knots with hand-measured proper motions.
Dashed line marks the estimated 1671 convergence date derived by \citet{Thor01}. 
}
\label{fig:knot_dates}
\end{figure}

\end{document}